\pdfoutput=1
\documentclass[10pt,twocolumn]{article}
\usepackage[a4paper, margin=1.5cm, bmargin=2cm]{geometry}
\usepackage{indentfirst}
\usepackage[utf8]{inputenc}
\usepackage[english]{babel}
\usepackage{amsmath}
\usepackage{amsfonts}
\usepackage{amssymb}
\usepackage[per-mode=symbol]{siunitx}
\usepackage{bm}                       
\usepackage{mathpazo}                 
\usepackage{microtype}                
\usepackage{tikz}
\usepackage{graphicx}
\usepackage{float}                    
\usepackage[export]{adjustbox}        
\usepackage[keeplastbox,nospread]{flushend}    
\usepackage[font=small,labelfont=bf]{caption}
\usepackage{subfig}
\usepackage{ifthen}                   
\usepackage{booktabs}                 
\usepackage{multirow}                 
\usepackage[symbol]{footmisc}         

\usepackage{hyperref}
\hypersetup{colorlinks=true, linkcolor=blue, citecolor=blue, urlcolor=blue} 

\usepackage{xcolor}
\definecolor{blue1}{RGB}{ 7,  47,  95}
\definecolor{blue2}{RGB}{18,  97, 160}
\definecolor{blue3}{RGB}{56, 149, 211}
\definecolor{red}{RGB}{210, 0, 0}

\usepackage{titlesec}                 
\titlelabel{\thetitle.\ }             
\titleformat*{\section}{\color{blue1}\scshape\bfseries\centering\large}
\titleformat*{\subsection}{\color{blue2}\normalfont\itshape\large}
\titleformat*{\subsubsection}{\color{blue3}\normalfont\itshape}
\titleformat{\paragraph}[runin]{\color{blue3}\normalfont\itshape}{}{0em}{}[~-]
\titlespacing{\paragraph}{0em}{0em}{0.3em}

\usepackage[backend=bibtex,date=year,doi=false,giveninits=true,isbn=false,maxnames=1,maxbibnames=99,sorting=none,style=nature]{biblatex}
\AtEveryBibitem{
	\clearlist{language}
}
\newlength{\spc} 
\let\footnoteorig\footnote
\renewcommand{\footnote}[2]{
	\ifthenelse{\equal{#2}{,}\OR\equal{#2}{.}}{%
		\settowidth{\spc}{#2}
		\addtolength{\spc}{-1.8\spc}
		#2
		\hspace*{\spc}
		\footnoteorig{#1}
	}{%
		\footnoteorig{#1}%
		\ifthenelse{\NOT\equal{#2}{;}\AND\NOT\equal{#2}{:}}{\ }{}%
		#2%
	}%
} 

\renewcommand{\textcite}[1]{\citeauthor{#1}\hspace*{-0.15em}\supercite{#1}}
\renewcommand{\cite}[2]{
	\ifthenelse{\equal{#2}{,}\OR\equal{#2}{.}}{%
		\settowidth{\spc}{#2}
		\addtolength{\spc}{-1.8\spc}
		#2
		\hspace*{\spc}
		\supercite{#1}
	}{%
		\supercite{#1}%
		\ifthenelse{\NOT\equal{#2}{;}\AND\NOT\equal{#2}{:}}{\ }{}%
		#2%
	}%
}

\newcommand{\snspace}[2][0.45em]{
	\hspace*{-#1}#2\hspace{0.2em}}

\begin{document}

\twocolumn[
	\begin{@twocolumnfalse}

		\begin{center}
			\textbf{\color{blue1}\Large An optimized material removal process}\\
			\vspace{1em}
			Jean-François Molinari\footnotemark\hspace*{-0.35em},\hspace{0.1em} Son Pham-Ba\\\vspace{0.5em}
			\textit{\footnotesize Institute of Civil Engineering, Institute of Materials Science and Engineering,\\\vspace{-0.2em}
			École polytechnique fédérale de Lausanne (EPFL), CH 1015 Lausanne, Switzerland}
		\end{center}


		\begin{center}
			\parbox{14cm}{\small
				\setlength\parindent{1em}We conduct boundary element simulations of a contact problem consisting of an elastic medium subject to tangential load. Using a particle swarm optimization algorithm, we find the optimal shape and location of the micro-contacts to maximize for a given load the stored elastic energy contributing to the removal of a spherical particle contained in between the micro-contacts. We propose an ice scream scoop as an application of this optimization process.
				
				\vspace{1em}
			}
		\end{center}

		\vspace{1em}

	\end{@twocolumnfalse}
]

\footnotetext{Corresponding author.\\E-mail address: \href{mailto:jean-francois.molinari@epfl.ch}{jean-francois.molinari@epfl.ch}}

\section{Introduction}

Wear, the process of material removal when two solids are in sliding contact, comes in various forms, adhesive and abrasive wear being the most prominent\cite{rabinowiczFrictionWearMaterials1995}. The formation of debris particles is often thought as a probabilistic event. It is known that natural or man made surfaces are rough over a range of length scales\cite{mandelbrotFractalCharacterFracture1984,majumdarFractalCharacterizationSimulation1990,thomNanoscaleRoughnessNatural2017}. It implies that the contact between nominally flat surfaces is in reality a contact between two rough surfaces when viewed microscopically, such that the real contact area is much smaller than the apparent contact area\cite{greenwoodContactNominallyFlat1966,bushElasticContactRough1975,perssonElastoplasticContactRandomly2001,hyunFiniteelementAnalysisContact2004}. Protruding asperities from both rough surfaces make junctions and result in what are called \emph{micro-contacts}. In the probabilistic view of wear, only a fraction of those
micro-contacts form debris particles.

Recent advances have permitted a leap forward on establishing a deterministic criterion for wear particle formation, at least in the context of adhesive wear. This new understanding emerged thanks to recent numerical studies performed at the small near-atomic scale\cite{aghababaeiCriticalLengthScale2016,aghababaeiDebrislevelOriginsAdhesive2017,brinkAdhesiveWearMechanisms2019}. The formation of wear particles at an unlubricated tribological interface due to adhesive wear was first theorized to be driven by a competition between deformation energy and fracture energy in 1958 by Rabinovicz\cite{rabinowiczEffectSizeLooseness1958}. This Griffith (fracture mechanics) approach to wear particle formation was recently extended to account for plastic flow, and validated with molecular dynamics (MD) simulations\cite{aghababaeiCriticalLengthScale2016}. The theory predicts the existence of a critical length scale $d^*$\snspace, dictating a transition between a ductile and a brittle behavior for a given material at a contact junction. Consequently, $d^*$ also corresponds to the minimal wear particle size which can be formed under adhesive wear when two asperities located on two opposed sliding surfaces collide into each other. $d^*$ was found to be mainly dependent on the material properties, with second order effects related to the geometry of the contacting asperities. While these works focused on adhesive wear, abrasive wear mechanisms can also be understood through the lens of fracture mechanics\cite{harishModelingTwobodyAbrasive2019}.

\begin{figure}
	\centering
	\subfloat[\label{subfig:agha-inter-a}]{
		\includegraphics[width=4cm]{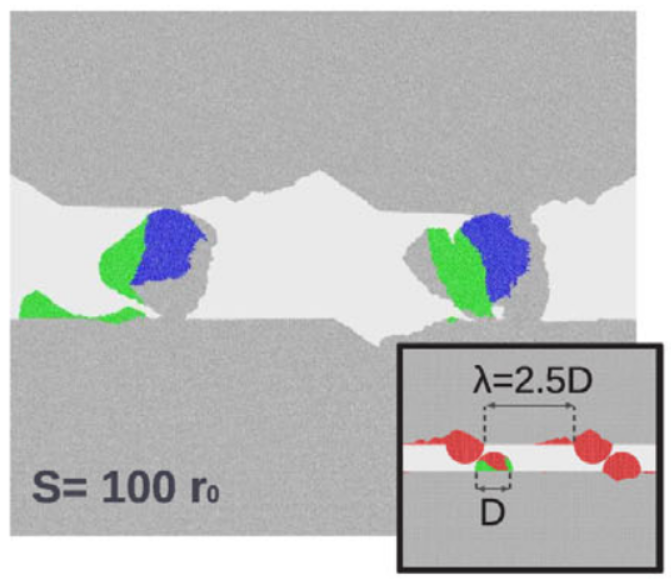}
	}
	\subfloat[\label{subfig:agha-inter-b}]{
		\includegraphics[width=4cm]{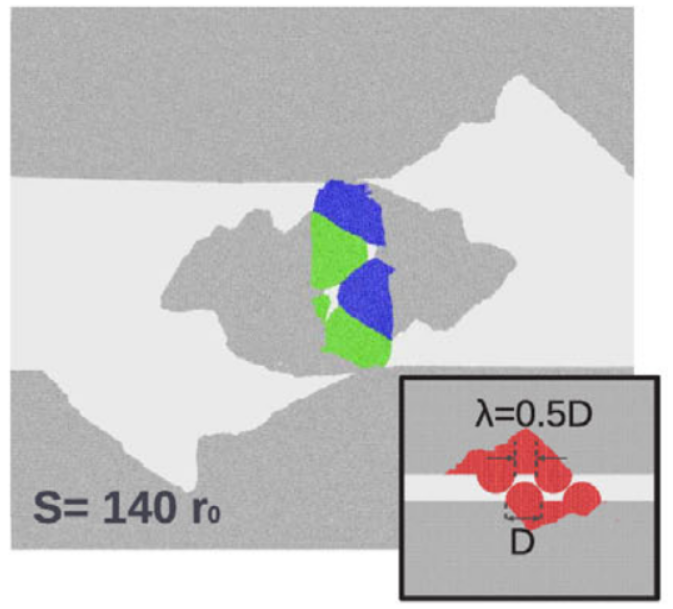}
	}
	\caption[2D MD simulations of asperity-level wear mechanisms.]{2D MD simulations of asperity-level wear mechanisms. The insets show the initial setups and positions of contact junctions. Atoms in red color end up in the formed fragments seen in the main figures when the system is sheared. This figure is reproduced from Aghababaei et al\cite{aghababaeiAsperityLevelOriginsTransition2018}. \textbf{(a)}~Contact junctions are initially far apart and each form an individual wear particle. \textbf{(b)} Contact junctions are close and, through elastic interactions, result in the formation of a single, much larger, wear particle.}
	\label{fig:agha-inter}
\end{figure}

Later, these numerical simulations were extended to account for interactions between nearby micro-contacts, each micro-contact being susceptible to result in the formation of a wear particle under the application of shear load\cite{aghababaeiAsperityLevelOriginsTransition2018} (see Fig.~\ref{fig:agha-inter}). Micro-contact junctions that are far from each other result in the formation of separated wear particles (Fig.~\ref{fig:agha-inter}\subref{subfig:agha-inter-a}). However, the simulations revealed that micro-contacts that are  close to each other, i.e. separated by a distance of the order (or less) than the junction size, result in the formation of a combined larger particle, due to crack shielding mechanisms (Fig.~\ref{fig:agha-inter}\subref{subfig:agha-inter-b}). This simple observation provides a mechanistic argument for the transition from mild to severe wear  observed at high loads, e.g. when the contact surface is populated by larger and denser micro-contacts thereby promoting elastic interactions between those. 
More recent theoretical considerations, supported by discrete MD simulations and simulations conducted in a continuum setting using the boundary element method, confirmed and extended these findings to multiple interacting junctions in a 2D setting\cite{pham-baAdhesiveWearInteraction2020}. Also noteworthy is the confirmation of the importance of crack shielding mechanisms for nearby contact junctions thanks to 2D finite-element simulations in which a phase-field formulation of fracture permitted a robust mesh-independent resolution of crack paths\cite{colletVariationalPhasefieldContinuum2020a}. The extent of such interactions remains to be thoroughly studied in 3D.

With the general understanding that elastic interactions between contact patches can increase the possibility of forming wear particles of a larger volume, this paper explores the uncharted territory of elastic interactions in a 3D setting. We aim to exploit those interactions by searching for an adhesive contact shape that maximises the volume of a detached chunk of material.  This shape would comprise of multiple adhesive regions, or in general, regions able to transmit a tangential load to the material to be carved. These tangential loads can be transmitted by a hard rigid tool indenting a soft elastic surface, thereby entering the realm of abrasive wear.  
Sect.~\ref{sec:wear} describes the wear criterion, which compares the adhesive energy required to create new surfaces to the stored elastic energy, evaluated using the boundary element method. Sect.~\ref{sec:opti} details the particle swarm optimization algorithm to probe contact patches shapes. Finally results are shown in Sect.~\ref{sec:results}. We propose as an application a novel ice cream scoop design.

\section{Micro-mechanics of wear}\label{sec:wear}

\subsection{Wear criteria}

We consider two surfaces that are sliding on each other, with the adhesive junctions formed between them being loaded tangentially\footnote{One can also consider a hard abrasive tool gripping an elastic body.}. The two bodies resist the sliding force, deforming, and accumulating elastic energy $E_\text{el}$. Considering one of the sliding bodies as a semi-infinite body, the elastic energy stored inside this body is
\begin{equation}\label{eq:Eel_integral}
	E_\text{el} = \frac{1}{2} \int_\Gamma \bm{u} \cdot \bm{p} \, d\Gamma \,,
\end{equation}
where $\Gamma$ is the nominally flat surface upon which load is applied, $\bm{p}$ is the traction field applied on this surface and $\bm{u}$ is the displacement field caused by the traction field.

At the scale of the contact junctions, wear is reduced to the formation of debris particles under the junctions. The detachment of a wear particle from a body requires the creation of new surfaces, thus requiring surface energy, or adhesive energy $E_\text{ad}$, which is proportional to the total surface area created times a surface energy $\gamma$.

When a wear particle is detached, the tangential load it was carrying can no longer be transmitted between the two sliding surfaces, resulting in a drop $\Delta E_\text{el}$ in the amount of stored elastic energy. Similarly to Griffith's criterion for crack propagation, a criterion can be established for the possibility to fully detach a wear particle of a given shape: the drop in elastic energy obtained when detaching the particle must be equal or greater than the amount of adhesive energy required:
\begin{equation}\label{eq:crit_E}
	\Delta E_\text{el} \geqslant E_\text{ad} \,.
\end{equation}

The other necessary condition for the detachment of a wear particle is to have a location where crack nucleation can occur. We assume that a crack can be initiated at a point if
\begin{equation}\label{eq:crit_s}
	\sigma_\text{I} \geqslant \sigma_\text{m} \,,
\end{equation}
where $\sigma_\text{I}$ is the first principal stress, or the maximum tensile stress if positive, and $\sigma_\text{m}$ is the tensile strength of the material.

Consequently, we are left with two criteria for wear particle formation: a \emph{crack initiation criterion} \eqref{eq:crit_s}, and an \emph{energetic feasibility criterion} \eqref{eq:crit_E}.

\subsection{Elastic energy computation}

\begin{figure}
	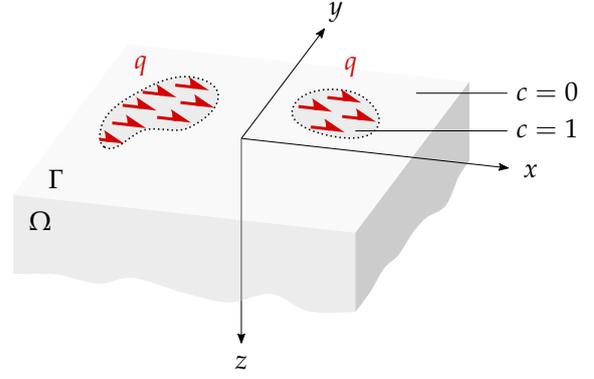
	\caption[Sheared adhesive junctions on a semi-infinite body]{Sheared adhesive junctions on a semi-infinite body. The function $c(x, y)$ describes the geometry of the junctions on the surface $\Gamma$.}
	\label{fig:setup}
\end{figure}

Let us consider a semi-infinite body $\Omega$ whose free surface $\Gamma$ is in the $(x, y)$ plane at $z = 0$. Some contact junctions are distributed on $\Gamma$ and are described by a `contact' field $c(x, y)$ equal to 1 where a junction is present and 0 otherwise. When a sliding force is applied on the body in the $x$ direction, we assume that the adhesive junctions will carry a uniform tangential stress $q$ also in the $x$ direction. No normal load is applied. The $x$ component of the surface traction field $\bm{p}$ on $\Gamma$ is therefore
\begin{equation}\label{eq:px}
	p_x(x, y) = c(x, y)q \,,
\end{equation}
with the other components in the $y$ and $z$ direction equal to 0. The setup is shown in Fig.~\ref{fig:setup}.

The surface displacements can be obtained from the surface tractions. A unit point load applied at the origin of $\Omega$ in the $x$ direction results in a displacement field whose $x$ component is\cite{johnsonContactMechanics1985a}
\begin{equation}\label{eq:ux_ker}
	u_{x\rightarrow x}^\text{ker} = \frac{1}{4\pi G} \left[ 2(1 - \nu)\frac{1}{r} + 2\nu\frac{x^2}{r^3} \right] \,,
\end{equation}
where $G$ is the shear modulus of the material, $\nu$ the Poisson's ratio, and $r$ is the distance from the origin: $r^2 = x^2 + y^2 + z^2$. There are also non-zero components of the displacement field in the $y$ and $z$ direction, but they are not relevant in this case, as shown below.

The surface displacements in the $x$ direction due to the full traction field \eqref{eq:px} is
\begin{align}
	u_x(x, y) &= \iint u_{x\rightarrow x}^\text{ker}(x - \xi, y - \eta) p_x(\xi, \eta) \, d\xi \, d\eta \\
	&= [u_{x\rightarrow x}^\text{ker} * p_x](x, y) \,, \label{eq:ux}
\end{align}
which is a convolution (denoted by the $*$ symbol). The expression for elastic energy \eqref{eq:Eel_integral} becomes
\begin{equation}\label{eq:Eel_x}
	E_\text{el} = \frac{1}{2} \int_\Gamma u_x p_x \, d\Gamma \,,
\end{equation}
where $u_x$ and $p_x$ are obtained from \eqref{eq:ux} and \eqref{eq:px}. Since $p_y = 0$ and $p_z = 0$, the components of displacement in those directions do not intervene in \eqref{eq:Eel_x}.

When dealing with this setup computationally, $\Gamma$ can be discretized into a finite grid, and the integral of \eqref{eq:Eel_x} can be turned into a finite sum. Care must be taken when considering a system of finite size, since the displacement kernel \eqref{eq:ux_ker} decreases when moving away from the origin but does not vanish before reaching infinity. Therefore, the contact region $c$ must not have non-zero values near the boundaries of the discretized finite $\Gamma$ to lower the impact of the finite size domain on the elastic energy computation. The discretization of the surface and the use of a kernel \eqref{eq:ux_ker} to compute the displacements from the surface tractions are part of the boundary element method (BEM)\cite{bonnetmarcBoundaryIntegralEquation1999}.

The computation of the elastic energy allows to check if the energetic feasibility criterion is satisfied, when the adhesive energy is already known (it is easily calculated from the estimated shape of the wear particle to be potentially formed). One particularity of choosing a traction distribution such as \eqref{eq:px} is that discontinuities in the function $c$ between 0 and 1 values cause stress singularities (regions of infinite stress). In reality, such stress singularities would be regularized, because materials get damaged or flow plastically above a certain stress. Nevertheless, those regions are likely to satisfy the crack initiation criterion, so we will assume that this criterion is always satisfied at the boundaries of the junctions defined by the function $c$.

For contact junctions that are far apart, individual wear particles can form beneath them, provided the energetic feasibility criterion is satisfied. When junctions are brought closer together, the elastic energy stored in the system increases due to elastic interactions\cite{pham-baAdhesiveWearInteraction2020,colletVariationalPhasefieldContinuum2020a}. This elastic energy increase can result in the formation of larger wear particles, encompassing multiple nearby junctions, as shown in Fig.~\ref{fig:agha-inter} in the 2D case.

\section{Material removal}\label{sec:opti}

\subsection{Problem statement}

We now wish to find the most efficient way to remove a piece of material from a body with a flat surface. We assume that the piece of removed material must have a roughly hemispherical shape of known diameter, so that $E_\text{ad}$ is also known and fixed. To minimize the effort put into the detachment of material, one must maximize $E_\text{el}$ by changing the shape of the contact junctions $c$ while trying to decrease the imposed tangential load, where the total tangential load is
\begin{equation}
	F_x = \int_\Gamma c q \, d\Gamma \,,
\end{equation}
which is deduced from \eqref{eq:px}.

The optimization problem is the following: find the function $c$, which maximizes $E_\text{el}$ for a given $F_x$ ($q$ is modified according to $c$ to keep $F_x$ constant).

\begin{figure}
	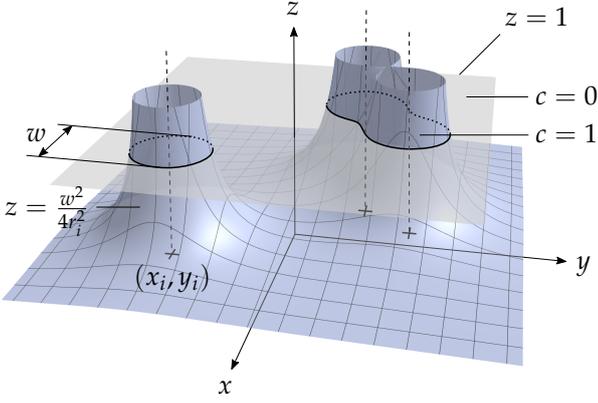
	\caption[Creation of metaballs from functions added together and clamped]{Creation of metaballs from functions added together and clamped. Each metaball adds a contribution to the plotted surface. The contact function $c$ is equal to $1$ whenever the summed surface is above $z=1$. On the left, an isolated metaball makes a circular shape in $c$. On the right, nearby metaballs are smoothly merged to construct the contact function $c$.}
	\label{fig:metaballs}
\end{figure}

To parameterize the function $c$, which is a binary representation of the shape of the sheared junctions, we use $n_\text{m}$ \emph{metaballs}\cite{blinnGeneralizationAlgebraicSurface1982}, which are n-dimensional\footnote{In the present case, they are two-dimensional.} circular objects usually used in computer graphics because of their organic appearance, as they smoothly merge with nearby metaballs. They allow us to create a complex shape $c$ using simple circular objects with smooth connections between them. Each metaball has a parameterized center $(x_i, y_i)$, with $1 \leqslant i \leqslant n_\text{m}$, and a fixed width $w$. A metaball adds a $w^2/(4r_i^2)$ term to $c$, where $r_i$ is the distance to its center: $r_i^2 = (x - x_i)^2 + (y - y_i)^2$. The whole function $c$ is the sum of all metaballs contributions, binarized to only keep regions where it is greater than 1. Mathematically:
\begin{equation}
	c(x, y) = \begin{cases}
		1 & \text{if } \displaystyle\sum_{i = 1}^n \frac{w^2}{4((x - x_i)^2 + (y - y_i)^2)} \geqslant 1 \,, \\
		0 & \text{otherwise.}
	\end{cases}
\end{equation}
This expression can be verified to work properly when a single metaball is present: it results in $c$ being non-zero in the region where $r_1 \leqslant w/2$, which is a circular region of width $w$ centered on $(x_1, y_1)$ and is the intended behavior. Fig.~\ref{fig:metaballs} illustrates how metaballs merge to create the function $c$.

\begin{figure}
	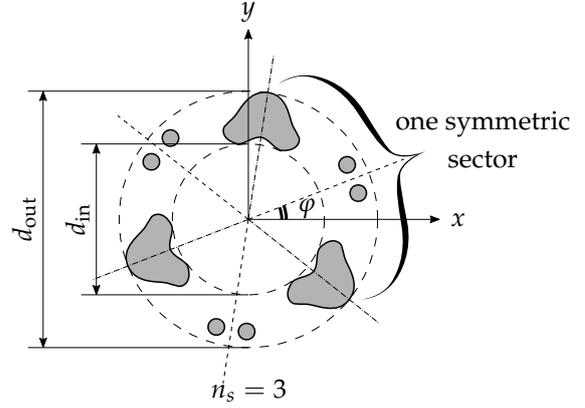
	\caption[Representation of the sectors of the contact function $c$]{Representation of the sectors of the contact function $c$. Each dark area is made of several metaballs. In this case, each of the three sectors has an imposed symmetry, which is not a mandatory  constraint. The inner diameter $d_\text{in}$ corresponds to the overall size of the piece of material to detach.}
	\label{fig:sectors}
\end{figure}

Geometrical constrains have to be put on the function $c$. It must have some edges coincident to the edge of the particle to be detached in order to satisfy the crack initiation criterion. Also, its overall size can be constrained to fit design limitations or to reduce the size of the search space, which has to be done carefully in order to maintain the performance of optimal solutions. To fit the geometrical constraints imposed on $c$, the metaballs are placed such that their centers are located in a ring of inner diameter $d_\text{in}$ and outer diameter $d_\text{out}$. The inner diameter corresponds to the size of the piece of material to detach, and the outer diameter limits the size of the search space of the contact zone, without loss of generality. Indeed, choosing a too large outer diameter would not yield a better optimized design because contact junctions that are far from each other do not interact elastically with each other. Another advantage of using a search space delimited by radii is that it brings natural symmetries. An $n_\text{s}$ fold symmetry can be imposed on the positions of the metaballs, and each sector can itself be symmetric. An orientation parameter $\varphi$ can be added to control the angular position of the axes of symmetry with respect to the direction of shear. Those sector parameters are represented in Fig.~\ref{fig:sectors}.

\subsection{Optimization}

The elastic energy has to be maximized by finding the optimal metaball parameters: the position of their centers, $x_i$ and $y_i$ for $1 \leqslant i \leqslant n_\text{m}$, and the global orientation $\varphi$. If all symmetry conditions are used (forcing $n_\text{s}$ identical and symmetric sectors), the total number of parameters is\footnote{There are $n_\text{m}$ parameterized metaballs with two coordinates, so $2n_\text{m}$ parameters. There are $n_\text{s}$ identical sectors, so this number is divided by $n_\text{s}$. Each sector is symmetric so this number is again divided by 2. The overall orientation adds 1.}
\begin{equation}\label{eq:n_param}
    n_\text{m}/n_\text{s} + 1 \,.
\end{equation}
The number of metaballs $n_\text{m}$ has to be taken large enough to have a fine control over the shape of $c$. To deal with the large number of parameters and the potential non-convexity of the problem, the choice of an evolutionary algorithm was opted for. Here, the \emph{particle swarm optimization} (PSO)\cite{kennedyParticleSwarmOptimization1995,zhangComprehensiveSurveyParticle2015} is used.

In PSO, a \emph{swarm} (a population) of \emph{particles} is considered. Each particle is a candidate solution (a shape of $c$) with a \emph{position} and a \emph{velocity}. The position\footnote{Not to be confused with the positions of the metaballs in $c$.} is the current set of parameters of the particle, and the velocity is the rate of change of each parameter between two iterations of the PSO. The position is bounded by the limits imposed on each parameter.

The swarm is initialized with $n_\text{p}$ particles having random initial positions within the bounds. Each solution is randomized such that the centers of its metaballs are uniformly distributed in the $(x, y)$ space (inside the ring of diameters $d_\text{in}$ and $d_\text{out}$). The positions were parameterized in polar coordinates $(r, \theta)$ to facilitate the enforcement of bounds. The initial velocities are also randomized, but such that an iteration of the PSO does not create an off-bound position (more details about the update process below).

\begin{figure*} 
    \centering
	\subfloat[\label{subfig:ring}]{
		\includegraphics[height=2.4cm, trim=1.5cm 1cm 3cm 0.5cm, clip]{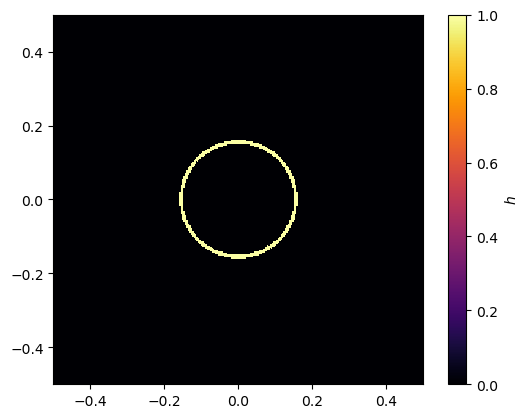}
	}
	\subfloat[\label{subfig:h1}]{
		\includegraphics[height=2.4cm, trim=1.5cm 1cm 3cm 0.5cm, clip]{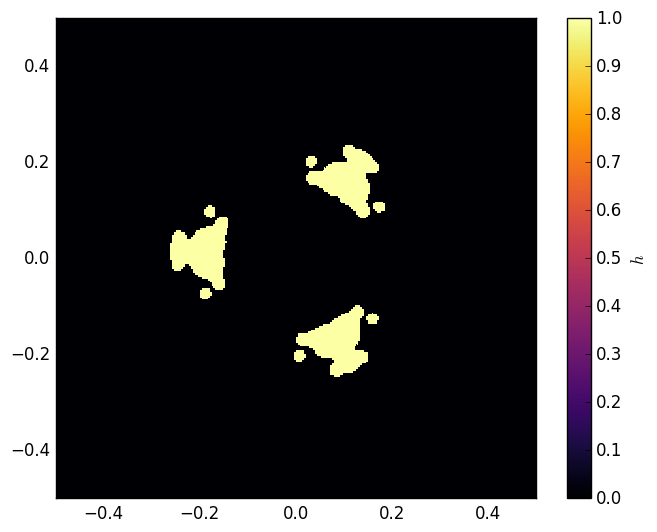}
	}
	\subfloat[\label{subfig:h2}]{
		\includegraphics[height=2.4cm, trim=1.5cm 1cm 3cm 0.5cm, clip]{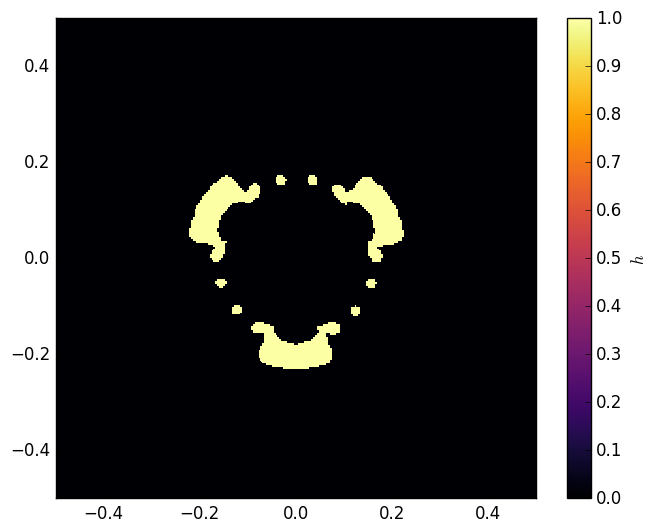}
	}
	\subfloat[\label{subfig:h3}]{
		\includegraphics[height=2.4cm, trim=1.5cm 1cm 3cm 0.5cm, clip]{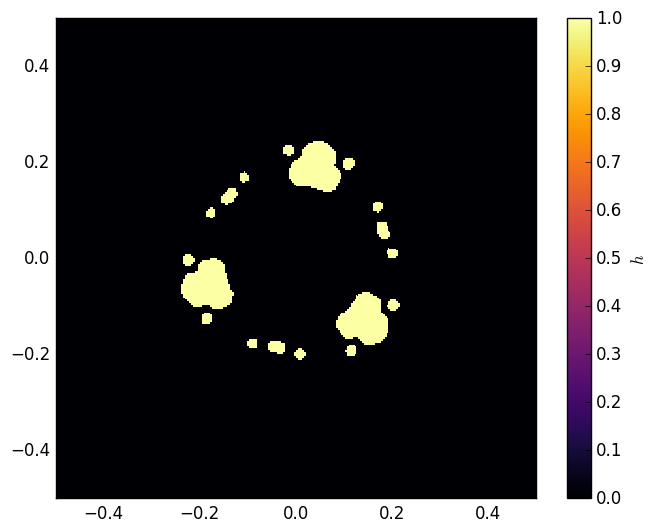}
	}
	\subfloat[\label{subfig:h4}]{
		\includegraphics[height=2.4cm, trim=1.5cm 1cm 3cm 0.5cm, clip]{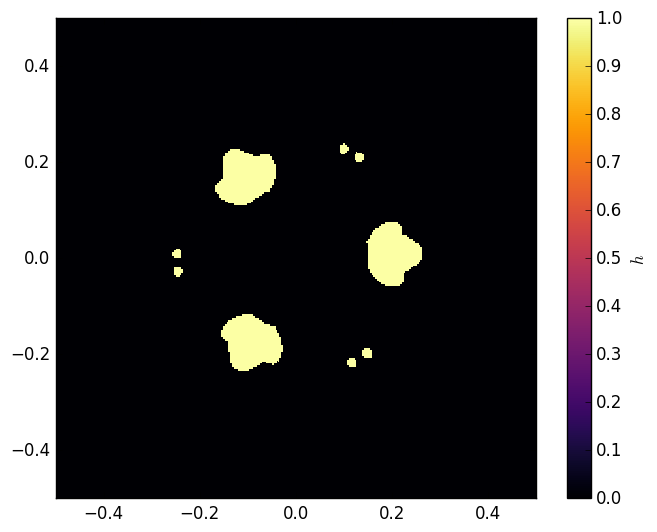}
	}
	\subfloat[\label{subfig:h5}]{
		\includegraphics[height=2.4cm, trim=1.5cm 1cm 3cm 0.5cm, clip]{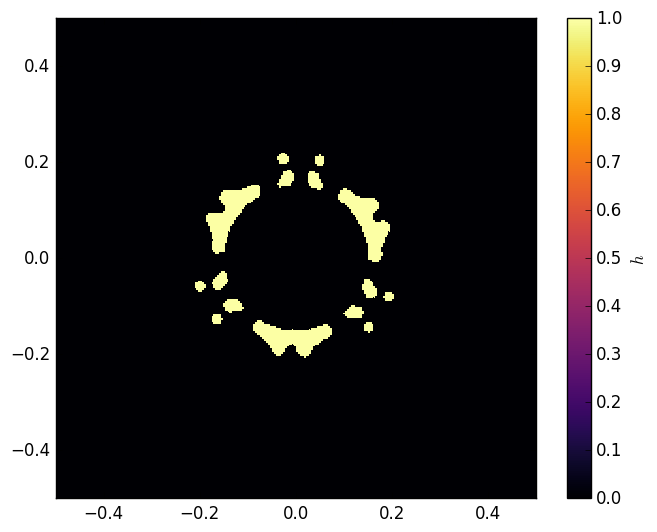}
	}
	\subfloat[\label{subfig:sol}]{
		\includegraphics[height=2.4cm, trim=1.5cm 1cm 3cm 0.5cm, clip]{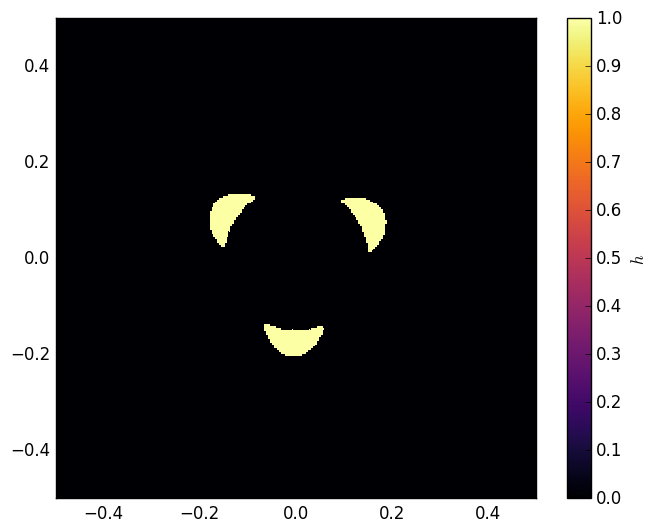}
	}
	\caption[Various shapes]{Various shapes. \textbf{(a)}~Trivial reference shape. \textbf{(b)-(f)}~Examples of random initial solutions with $n_\text{m} = 60$ metaballs of width $w = 0.019$ and $n_\text{s} = 3$ identical and symmetrical sectors. \textbf{(g)}~Optimized shape.}
	\label{fig:shapes}
\end{figure*}

At each iteration $t$, the particles compute their objective function (the value of stored elastic energy) at their current position (in the space of all parameters) $p_j^t = [\varphi, r_1, \theta_1, r_2, \theta_2, ...]$ and update their memory of best visited position $p_j^\text{b}$ if necessary. The particles have inertia, so their velocity $v_j^t$ is conserved up to a factor $k_v \leqslant 1$. The particles are also attracted toward their own best visited position $p_j^\text{b}$ and toward the overall best position visited by the swarm $p^\text{B}$\snspace, which have an influence on the particles velocity thanks to the hyperparameters $k_\text{b} \leqslant 1$ and $k_\text{B} \leqslant 1$. To summarize, the velocity of each particle is updated as follow:
\begin{equation}
	v_j^t = k_vv_j^{t-1} + k_\text{b}(p_j^\text{b} - p_j^t) + k_\text{B}(p^\text{B} - p_j^t)
\end{equation}
and their position is updated as
\begin{equation}
	p_j^t = p_j^{t-1} + v_j^t \ \Delta t
\end{equation}
with $\Delta t = 1$. The positions $p_j^t$ are kept inside their bounds after each update by clamping their metaballs polar coordinates $r_i$ inside a ring and $\theta_i$ inside a sector (see Fig.~\ref{fig:sectors}). $k_v$, $k_\text{b}$ and $k_\text{B}$ are hyperparameters of the optimization method and have to be fixed by the user of the method.

\section{Results}\label{sec:results}

\subsection{Optimal shape}

Since the problem is a matter of maximizing the elastic energy while keeping other dimensions (such as tangential load and maximum overall size) constant, we can work with adimensionalized unitless quantities.

We use a discretized space of size $1 \times 1$ and with a resolution of $256 \times 256$ for the computation of $E_\text{el}$ from $c$. The Young's modulus of the material is set to $E=1$ and its Poisson's ratio to $\nu = 0.3$. The tangential load is set to $F_x = 0.3$ (its choice has no incidence on the results).

The whole function $c$ is made of a total of $n_\text{m} = 60$ metaballs of width $w = 0.019$ and has $n_\text{s} = 3$ identical and symmetrical sectors. With those symmetry conditions on the metaballs, there are 10 independent metaballs per half-sector and therefore 21 parameters per solution according to \eqref{eq:n_param}. The centers of the metaballs are constrained in a ring of diameters $d_\text{in} = 0.3$ and $d_\text{out} = 0.5$. Fig.~\ref{fig:shapes} (\subref{subfig:h1} to \subref{subfig:h5}) shows examples of initial positions (solutions).

For the PSO, $n_\text{p} = 200$ particles are used, with the hyperparameters $k_v = 0.97$, $k_\text{b} = 0.2$ and $k_\text{B} = 0.2$. The optimization is run for 50 iterations using a custom code, resulting in the solution shown in Fig.~\ref{fig:shapes}\subref{subfig:sol}. The optimization routine was performed five times with different randomized initial particles to ensure that it was not stuck in local optima.

In the following text, we express the objective function of a particle as its elastic energy $E_\text{el}$ divided by the elastic energy $E_{\text{el}, 0}$ of a trivial shape fulfilling the crack initiation criterion, i.e. a thin ring of diameter $d_\text{in}$ (see Fig.~\ref{fig:shapes}\subref{subfig:ring}). In the considered discretized space, we have $E_{\text{el}, 0} = 12700$.

On average, the objective function $E_\text{el} / E_{\text{el}, 0}$ of random initial solutions evaluates to 0.71. The optimized shape (Fig.~\ref{fig:shapes}\subref{subfig:sol}) has an objective function of 1.21, meaning it is 20\% more efficient energetically than the thin ring. Therefore, we have an increased performance compared to trivial shapes. The overall orientation $\varphi$ has a negligible influence on the results.

In order to check for convergence, a finer description of $c$ is used, with $n_\text{m} = 120$ and $w = 0.013$ to keep the overall surface area of $c$ constant. The optimization is run for 100 steps, and results in a best objective function at 1.24 and visually indistinguishable shapes compared to $n_\text{m} = 60$.

\begin{figure}
	\input{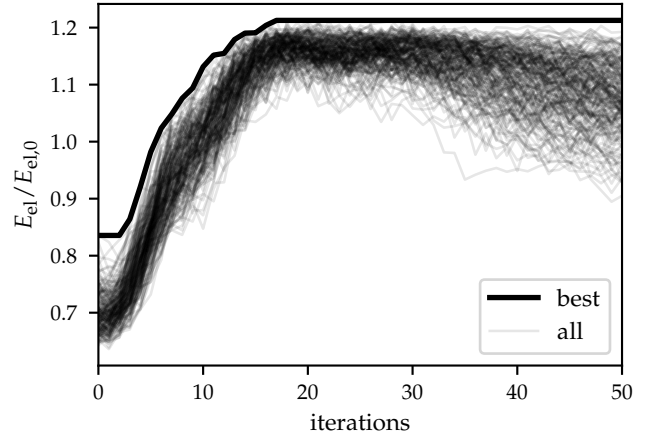}
	\caption[Evaluations of the objective function for each particle during the optimization]{Evaluations of the objective function for each particle during the optimization. The vertical axis shows the objective function relative to the evaluation of the shape Fig.~\ref{fig:shapes}\subref{subfig:ring}. The thickest line is the overall best at a given iteration, going from the random shape Fig.~\ref{fig:shapes}\subref{subfig:h1} to the optimized shape Fig.~\ref{fig:shapes}\subref{subfig:sol}.}
	\label{fig:sol_conv}
\end{figure}

\subsection{Example of application: ice cream scoop}

\begin{figure*}
	\centering
	\subfloat[Isometric view\label{subfig:iso}]{
		\includegraphics[width=5cm]{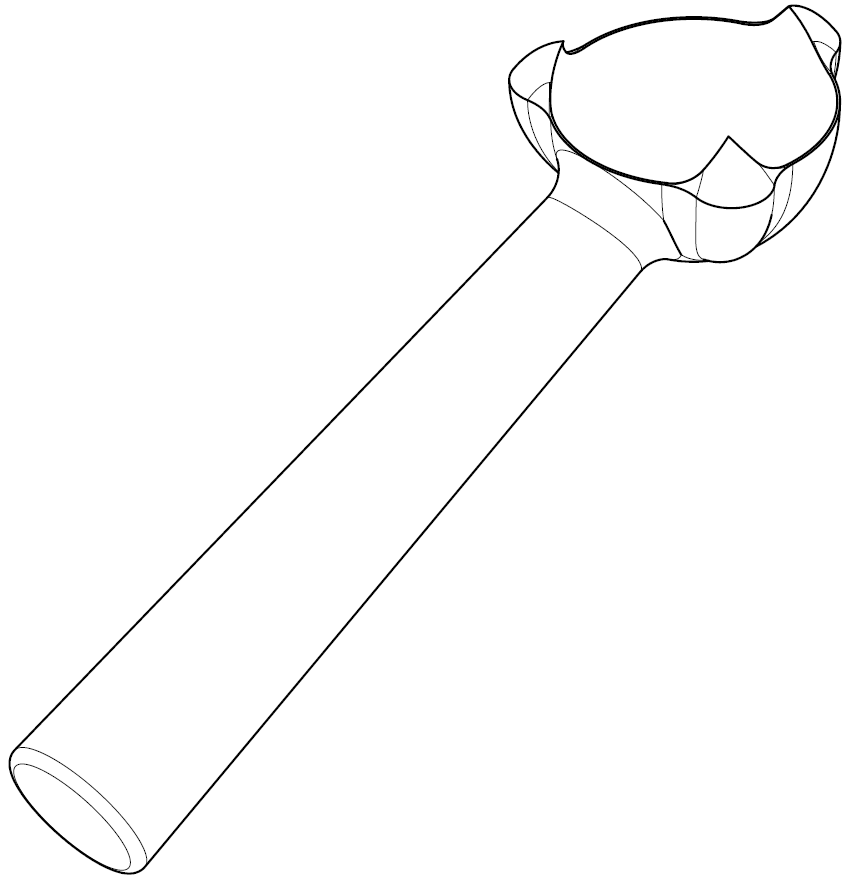}
	}
	\hspace{1.7cm}
	\subfloat[Cross section of the claws\label{subfig:section}]{
		\includegraphics[width=5cm]{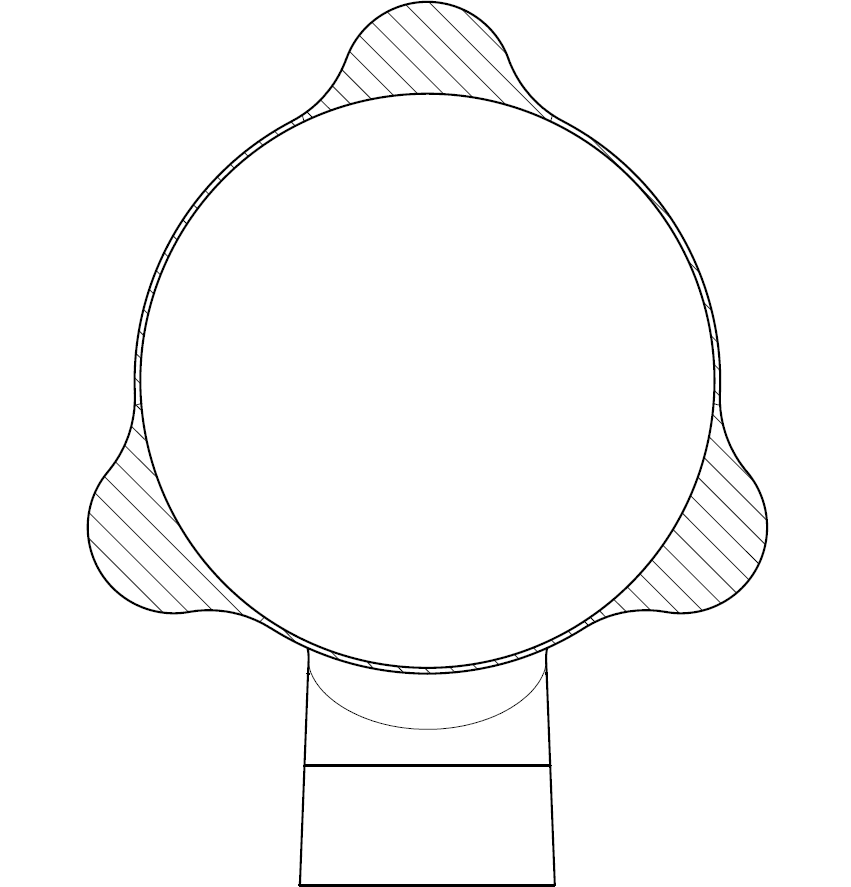}
	}
	\caption[Design of optimized ice scoop]{Design of optimized ice scoop. \textbf{(b)}~The cross section of the claws matches the optimized sheared shape.}
	\label{fig:scoop}
\end{figure*}

One practical way to use the newly found optimal contact shape is to design a tool for ice cream scooping. The size of the ball to create is fixed, and the amount of force needed to detach it can be minimized by utilizing the optimized shape. When the tool is used, it must pull on the surface of the ice cream while having the same contact shape as the one shown in Fig.~\ref{fig:shapes}\subref{subfig:sol}, instead of the usual one shown in Fig.~\ref{fig:shapes}\subref{subfig:ring}. To this end, claws can be added on a basic hemispherical scoop design, that will penetrate into the surface of the ice cream and create the desired pulling pattern. Fig.~\ref{fig:scoop}\subref{subfig:iso} shows a simple design implementing this idea of penetrating claws, and Fig.~\ref{fig:scoop}\subref{subfig:section} shows a cross-sectional view of the pattern formed by the penetration of the claws into the surface of the ice cream, matching the optimized shape found above. This design and the optimization method are patented\cite{molinariShovellingToolMethod2020}.

\section{Conclusion}

We have explored the design space of contact junctions location and shape to maximize the energetic efficiency of material removal, thanks to elastic interactions. This study extends to a three-dimensional setting previous efforts that were limited to two dimensions, thereby providing a much richer design space. The numerical approach combines the efficient boundary element method to solve the contact problem to a particle swarm optimization algorithm to search the optimal location and shape of contact junctions. We have found a three claws design that increases the energetic  efficiency of 20 percent. We propose an ice scream scoop application for which we have filed a patent.

\printbibliography

\end{document}